# RDD-Eclat: Approaches to Parallelize Eclat Algorithm on Spark RDD Framework
## (Extended Version)

Pankaj Singh[a 1], Sudhakar Singh[b 2 *], P K Mishra[b 3], Rakhi Garg[c 4]

[a]*Faculty of Education, Banaras Hindu University, Varanasi, India*
[b]*Department of Electronics and Communication, University of Allahabad, Prayagraj, India*
[c]*Mahila Maha Vidyalaya, Banaras Hindu University, Varanasi, India*
[1]psingh.edu@bhu.ac.in, [2]sudhakar@allduniv.ac.in, [3]mishra@bhu.ac.in, [4]rgarg@bhu.ac.in
[*]*Corresponding Author*

**Abstract**

Frequent itemset mining (FIM) is a highly computational and data intensive algorithm. Therefore, parallel and distributed FIM algorithms have been designed to process large volume of data in a reduced time. Recently, a number of FIM algorithms have been designed on Hadoop MapReduce, a distributed big data processing framework. But, due to heavy disk I/O, MapReduce is found to be inefficient for the highly iterative FIM algorithms. Therefore, Spark, a more efficient distributed data processing framework, has been developed with in-memory computation and resilient distributed dataset (RDD) features to support the iterative algorithms. On this framework, Apriori and FP-Growth based FIM algorithms have been designed on the Spark RDD framework, but Eclat-based algorithm has not been explored yet. In this paper, RDD-Eclat, a parallel Eclat algorithm on the Spark RDD framework is proposed with its five variants. The proposed algorithms are evaluated on the various benchmark datasets, and the experimental results show that RDD-Eclat outperforms the Spark-based Apriori by many times. Also, the experimental results show the scalability of the proposed algorithms on increasing the number of cores and size of the dataset.

## 1. Introduction

The revolution in technology for storing and processing big data leads to data intensive computing as a new paradigm. It requires efficient and scalable data mining techniques to find the valuable and precise knowledge from the big data. Data mining is the process of extracting hidden and interesting patterns from the huge volume of data [1]. In data mining, different types of techniques are applied depending on the kind of knowledge to be mined. To discover the interesting correlations among data objects of database, association rule mining (ARM) technique [2] of data mining is employed. Association rules are generated from the frequent itemsets which are computed by frequent itemset mining (FIM) algorithms. Apriori [2], Eclat [3], and FP-Growth [4] are the three basic algorithms of frequent itemset mining. A number of their variants and extensions have been developed for efficient mining of the frequent itemsets. These are the sequential algorithms executing on the single machine that is limited by the computing and memory capacity. So, these algorithms have been parallelized to be executed on the parallel and distributed computing systems [5]. But, these parallel and distributed algorithms are not able to handle and process big data efficiently, as the traditional distributed systems is based on

---

exchanging the data that requires high communication and network bandwidth. Further, it is lacking of fault tolerance, and high level parallel programming language. A distributed system that moves the computation to where the data is, rather than moving the data, has been developed named Hadoop [6]. Hadoop is a fault tolerant large-scale distributed batch processing system for big data. Its two core components are HDFS (Hadoop Distributed File System) [6] and MapReduce [7]. MapReduce is a scalable and parallel programming model that enables parallel processing of data resided in HDFS. HDFS stores big data in the form of blocks on the different nodes of Hadoop cluster, and a MapReduce job is executed as multiple independent tasks on the different splits of data across the nodes of Hadoop cluster.

MapReduce is scalable and fault tolerant, but not efficient for the iterative algorithms of data mining due to high I/O and network overhead in writing/reading intermediate results in HDFS. Further, invoking a new MapReduce job each time for a new iteration is time consuming that increases the overall execution time of algorithms. Spark [8-9], a more efficient and faster big data processing platform, overcomes the problems with Hadoop MapReduce, and is augmented with a number of powerful features like Resilient Distributed Datasets (RDDs) [10], rich set of operations including map and reduce, in-memory computation, and support to batch, interactive, iterative and streaming processing of data. RDD is a collection of immutable data objects partitioned across the nodes of Spark cluster. Spark stores the intermediate results in memory that reduces the number of read/write cycles to the storage system. It is 100 times faster in memory and 10 times faster on disk than Hadoop MapReduce [8]. Many authors have designed different FIM algorithms on the Spark RDD framework [11-17], in which most of the algorithms follow Apriori as the base algorithm. Parallelization of Eclat-based algorithm on Spark has not been explored yet to the best of our knowledge. In this paper, we consider Eclat [3], a more efficient algorithm than Apriori. Eclat reduces I/O cost due to a small number of database scan, and computation cost due to vertical data format and lattice traversal scheme.

This paper proposes some approaches to parallelize Eclat algorithm on the Spark RDD framework. The name RDD-Eclat represents Spark-based Eclat algorithm, and we propose five variants of RDD-Eclat named in short as EclatV1, EclatV2, EclatV3, EclatV4, and EclatV5. EclatV1 is the first version of the algorithm, and each subsequent version results from the further modifications on the preceding version to achieve better performance. Algorithm EclatV1 first generate frequent items and a vertical dataset. From vertical dataset, it constructs equivalence classes based on common 1-length prefix. A default partitioner partitions the equivalence classes into *(n-1)* independent partitions, where *n* is the number of frequent items. Equivalence classes in each partition are processed in parallel by applying the bottom-up search recursively on each equivalence class to enumerate the frequent itemsets. EclatV2 applies all operations of the algorithm on the filtered transactions which contain transactions with only frequent items. Transaction filtering is adopted from the efficient implementation of Apriori and Eclat by Borgelt [18]. EclatV3 is slightly different from EclatV2, and the difference is the use of accumulator, a kind of shared variable in Spark. Algorithms EclatV4 and EclatV5 are similar to EclatV3 except the partitioner used to partition the equivalence classes. These two algorithms use two different types of hash partitioner to partition the equivalence classes into *p* independent partitions, where *p* is the user defined value. The performance of our proposed algorithms is compared with the Spark-based Apriori algorithm on both synthetic and real life datasets, and they significantly outperform the Spark-based Apriori in terms of execution time. Further, the performance of all proposed RDD-Eclat algorithms is compared with each other in terms of speed and scalability to study the effect of various strategies applied on these algorithms.

The major contributions of this paper can be listed as follows:

1. Parallel Eclat algorithm on the Spark RDD framework is proposed named as RDD-Eclat.
2. RDD-Eclat is implemented as five different variants named as EclatV1, EclatV2, EclatV3, EclatV4, and EclatV5. These variants are based on the different strategies applied in speculation that better performance will be achieved.
3. Transaction filtering is adopted on some variants of algorithm to observe the effect on space and time complexity.
4. Heuristics for partitioning of equivalence classes are applied on some variants of algorithms to achieve partitions with balanced workload.
5. Extensive experiments are carried to compare the performance of all proposed algorithm with Spark-based Apriori. The proposed algorithms are also closely compared with each other in terms of speed and scalability.

The rest of the paper is organized as follows. Section 2 presents preliminaries for RDD-Eclat, which is a brief description of frequent itemset mining, Eclat algorithm, and Apache Spark. Section 3 discusses the related work. In section 4, the proposed algorithms are described in detail. Experimental results and analysis are presented in section 5. Finally, section 6 concludes the paper with future directions.

## 2. Preliminaries

### 2.1 Frequent Itemset Mining and Eclat Algorithm

Let $I = \{i_1, i_2, ..., i_m\}$ be a set of $m$ distinct *items* or attributes. An *itemset* is a set of items from the set $I$. A *k-itemset* is an itemset consisting of $k$ items. An itemset $X$ is said to be a subset of an itemset $Y$, denoted as $X \subseteq Y$, if all items of $X$ are also in $Y$. Let $D = \{T_1, T_2, ..., T_n\}$ be the database of $n$ transactions, such that each transaction $T_i$ comprise of an unique transaction identifier $TID_i$ and an itemset, i.e. in the form of $<TID_i, i_1, i_2, ..., i_k>$. A transaction $T_i$ contains an itemset $X$ if $X \subseteq T_i$. The number of transactions in $D$, containing the itemset $X$ is called *support count* or occurrence frequency of $X$, denoted as $\sigma(X)$. An itemset $X$ is said to be frequent if $\sigma(X) \geq min\_sup$, where $min\_sup$ is a user-specified minimum support threshold. Frequent itemset mining is the computation of all frequent itemsets in a given database [2]. Association rule mining [2] is a two step process. The first step generates all frequent itemsets, and is computationally intensive. The second step simply produces all confident association rules. An association rule is a conditional implication of the form $X => Y$, where $X, Y \subset I$, and $X \cap Y = \phi$. The confidence of the rule $X => Y$ is measured as $\sigma(X \cup Y) / \sigma(X)$, in terms of support count of itemsets. A rule is confident if its confidence is more than $min\_conf$, a user-specified minimum confidence threshold. The generation of all frequent itemsets is a computationally as well as memory, and disk I/O intensive task [19]. Various efficient algorithms have been developed to overcome these issues.

Eclat algorithm [3] uses a vertical tidset database format, equivalence class clustering, and bottom-up lattice traversal; consequently, reduces I/O and computation cost, as well as memory requirement. Eclat converts horizontal database into vertical database, i.e. from itemset format $<TID_i, i_1, i_2, ..., i_k>$ to tidset format $<i_k, TID_1, TID_2, ..., TID_k>$. A vertical tidset database consists of a list of items followed by respective tidsets. The tidset of an item or itemset $X$ is the set of all transaction identifiers containing $X$, and is denoted as $tidset(X) = \{T_i.TID \mid T_i \in D, X \subseteq T_i\}$. The support of an item or itemset $X$ is the number of elements in $tidset(X)$ i.e. $\sigma(X) = |tideset(X)|$ [20]. The tidset approach reduces the cost of support counting. The support of a candidate k-itemset is computed by the intersection of tidsets of its two (k-1)-subsets. The vertical database is more compact than horizontal and contains all relevant information, which reduces memory requirements and scanning of the whole database. Further, as the length of itemsets increases, their tidset decrease, that consequently reduces the cost of intersection operations. The computation of frequent 2-itemsets is costlier with vertical format in comparison to the horizontal format. So a triangular matrix is used to update the counts of candidate 2-itemsets [3] [19].

The set of items $I$ of the database forms a power-set lattice $\rho(I)$. The set of atoms of this lattice corresponds to the set of items [3]. The power-set lattice is the search space that contains all the potential frequent itemsets. To enumerate all the frequent itemsets, lattice must be traversed along with intersection operations on tidsets to obtain support count of itemsets. The equivalence class clustering partitions the lattice into smaller independent sub-lattices enabling parallel computation of frequent itemsets. It also overcomes the limited memory constraint when the complete lattice could not fit in memory due to the large intermediate tidsets. The equivalence class clustering partitions the itemsets of lattice into *equivalence classes* based on the common prefixes of itemsets. Suppose, the set of frequent k-itemsets $L_k$ is lexicographically sorted, then its itemsets can be partitioned into equivalence classes based on their common *(k-1)* length prefixes. For example, let $I = \{1, 2, 3, 4, 5\}$ and after preprocessing, the frequent 2-itemsets $L_2 = \{12, 13, 14, 15, 23, 24, 25, 34, 35, 45\}$. The prefix based equivalence classes will be as: *[1] = {2, 3, 4, 5}; [2] = {3, 4, 5}; [3] = {4, 5}; [4] = {5}*, where *[i]* represents the class identifier with prefix *i*. The candidate (k+1)-itemsets are generated by joining all pairs within an equivalence class, together with the prefix. So, the 3-itemsets produced by each class will be *{123, 124, 125, 134, 135, 145}; {234, 235, 245}; {345}* respectively [19]. Each of the k-length prefix based equivalence classes form a sublattice. For example, *[1]* is a lattice with atoms *{12, 13, 14, 15}*. All the classes can be processed independently and parallely, and if a class is large enough to be solved in main memory, it can be decomposed to the next level. Eclat uses a bottom-up lattice traversal scheme [3] that processes each equivalence class by recursively decomposing into smaller classes to enumerate all frequent itemsets. The pseudo code in the Algorithm 1 shows the recursive procedure of this bottom-up search technique, originally given by Zaki [3]. Here, $L_{EC_k}$ represents the set of frequent itemsets generated by the equivalence class $EC_k$.

**Algorithm 1:** Bottom-Up recursive function of Eclat

Input: $EC_k = \{A_1, A_2, ..., A_n\}$, equivalence class of k-itemsets consists of atoms $A_i$.
Output: Frequent itemsets $\in EC_k$

```
1:   Bottom-Up(EC_k)
2:   {
3:      for(i = 1; i <= |EC_k|; i++)
4:      {
5:         EC_{k+1} = ϕ;
6:         for(j = i + 1; j <= |EC_k|; j++)
7:         {
8:            A_{i j} = A_i U A_j;
9:            tidset(A_{i j}) = tidset(A_i) ∩ tidset(A_j);
10:           if( |tidset(A_{i j})| >= min_sup)
11:           {
12:              EC_{k+1} = EC_{k+1} U A_{i j};
13:              L_{EC_k} = L_{EC_k} U A_{i j};
14:           }
15:        }
16:        if(EC_{k+1} != ϕ)
17:           Bottom-Up(EC_{k+1});
18:     }
19:     return L_{EC_k};
20:  }
```

## 2.2 Apache Spark

Apache Spark [8] is a fast and general cluster computing system for large-scale batch and streaming data processing, originally developed at AMPLab of UC Berkeley [9-10]. Spark was developed to overcome the inefficiency of Hadoop MapReduce [7] [21] for iterative jobs and interactive analytics. It retains the good properties of MapReduce like scalability and fault tolerance. The backbone of Spark is a distributed memory abstraction called Resilient Distributed Datasets (RDDs) [10], which is a collection of immutable data objects partitioned across the nodes of Spark cluster. Spark achieves fault tolerance through a lineage chain that keeps record of set of dependencies on parent RDDs i.e. how an RDD derived from another RDD. A lost partition of RDD can be rebuilt quickly through the lineage chain.

A Spark application has a driver program running the main() function that implements the control flow of application and launches different operations on the Spark cluster in parallel. A Spark cluster consists of a master node and a number of worker nodes; all can be on a single machine. The architectural details can be found in [22]. The driver program manages a number of executors that are the processes launched for an application on the worker nodes. Executor runs tasks and keeps data in memory or disk storage. A task is a unit of work sent to an executor.

The RDDs can be created in four ways: by parallelizing an existing collection in the driver program, from an external storage system like HDFS, by transformation of exiting RDDs, and by changing the persistence of existing RDDs [9]. The RDDs provide two types of operations: transformations and actions. The transformation operations (e.g. map, flatMap, filter, reduceByKey etc.) create a new RDD from an existing one, while the action operations (e.g. collect, count, saveAsTextFile etc.) return the result to driver program or writes to external storage, after applying the relevant computation on RDD. Spark uses lazy evaluation technique, i.e. the transformations will not execute until an action is triggered [23]. Spark also provides two restricted types of shared variables: broadcast variables and accumulators. The broadcast variable is a mechanism to efficiently distribute a copy of a large read-only data to every worker to be cached on. The accumulators are variables that workers can only "add" to through an associative and commutative operation, and only the driver can read it [9].

## 3. Related Work

There is a long list of algorithms for the frequent itemset mining, proposed and developed by a number of researchers. These FIM algorithms can be classified into four major categories on the basis of computing paradigms. These paradigms are Sequential, MPI (Message Passing Interface), Hadoop MapReduce, and Spark. Sequential algorithms are single-machine based algorithms, for example, the three well known algorithms: Apriori [2], Eclat [3], and FP-Growth [4]. A number of improved sequential algorithms were proposed based on these representative algorithms. Most of the sequential algorithms were parallelized on parallel and distributed computing system using MPI to overcome the constraint of memory and computing speed of a single machine [5]. With the evolution of big data, re-designing of traditional data mining algorithms on Hadoop and Spark have been started to provide the scalability. A comprehensive survey of FIM algorithms on various computing paradigms can be found in [24].

With the introduction of Hadoop, researchers have proposed several FIM algorithms on Hadoop MapReduce framework based on the central algorithms Apriori, Eclat, and FP-Growth. The effort on the Apriori-based algorithm is more in comparison to Eclat and FP-Growth. Lin et al. [25] proposed three variants of MapReduce-based Apriori named as SPC (Single Pass Counting), FPC (Fixed Passes Combined-counting), and DPC (Dynamic Passes Combined-counting). SPC is a simple straight forward implementation of Apriori on MapReduce framework very similar to the algorithms proposed by other researchers [26-27]. FPC combines three consecutive passes (k, k+1, k+2; for k > 2) of candidate generation in a single MapReduce phase. DPC overcomes the problem of overloaded phase in FPC, so combines consecutive passes of candidate generation dynamically. Combining multiple passes in a single MapReduce phase minimizes the number of MapReduce phases that significantly reduces the overall execution time. DPC determines the number of passes to be combined based on the execution time of the preceding phase. But, this execution time may vary on the different datasets, as well as on the computing power of Hadoop cluster. Singh et al. [28] resolved the problems with FPC and DPC, and proposed two robust and proper algorithms named as VFPC (Variable Size based Fixed Passes Combined-counting), ETDPC (Elapsed Time based Dynamic Passes Combined-counting). They further optimized these algorithms by skipping pruning steps in some passes, and proposed another pair of algorithms named as Optimized-VFPC and Optimized-ETDPC [28]. Kovacs and Illes [29] applied the combiner functionality inside the Mapper and candidate generation inside the Reducer unlikely to the other algorithms. They also used a technique to count 1 and 2–itemsets in one step with a triangular matrix. PARMA, a parallel randomized algorithm on MapReduce framework is proposed by Riondato et al. [30] for discovering approximate frequent itemsets. In this algorithm, various machines of the Hadoop cluster process random samples of database. Jen et al. [31] tried to reduce the computation cost of MapReduce-based Apriori using a vertical data layout. Singh et al. [32] have investigated the effect of data structures on Apriori algorithm in MapReduce context. Recently, Chon and Kim [33] proposed BIGMiner, an Apriori-based frequent itemset mining algorithm on MapReduce. The authors have focused the three problems in order to design the algorithms: workload skewness, generation of large intermediate data during mining, and large overhead of network communication. Algorithm consists of two phases: pre-computation and frequent itemsets generation. The pre-computation phase generates transactions in bitmaps called transaction chunk for a specific length frequent itemsets. The second phase repeats candidate generation and testing steps of Apriori.

Two distributed versions of Eclat algorithms on MapReduce have been proposed by Moens et al. [34]. The algorithms are named as Dist-Eclat and BigFIM. Dist-Eclat partitions the search space on Mappers rather than data space. Big-FIM is a hybrid of Apriori and Eclat approaches. Initially, MapReduce-based Apriori approach is used to find frequent itemsets up to a specific length. After that, each Mapper applies Eclat algorithm independently on equivalence classes constructed from frequent itemset generated using Apriori approach. Liu et al. [20] have incorporated three improvements in Eclat algorithm and proposed Peclat (Parallel Eclat) algorithm that parallelizes this improved algorithm on MapReduce framework. Peclat uses a hybrid vertical format, mixset that automatically select tidset or diffset for each itemset. It applies Apriori's pruning to avoid unnecessary intersections of tidsets, and arrange itemsets in ascending order of support to balance the workloads.

PFP (Parallel FP-Growth) [35] is a MapReduce-based FP-Growth algorithm. It breaks the FP-Tree into smaller independent FP-Trees, which are processed by different Mappers to generate frequent itemsets. Zhou et al. [36] proposed BPFP, a balanced version of PFP algorithm. BPFP balances the groups of FP-Trees based on the frequencies of frequent items. FiDoop [37], a parallel frequent itemset mining algorithm on MapReduce uses an FIU-Tree (frequent items ultrametric tree) in the place of FP-Tree. The further extension of this algorithm called FiDoop-HD has also been proposed by Xun et al. [37] for the processing of high dimensional data.

The development of Spark has shifted the research focus from Hadoop MapReduce-based algorithms to the Spark-based algorithms. MapReduce does not fit in with the iterative nature of data mining algorithms. Each time, for a new iteration, one needs to launch a new MapReduce job that takes a significant amount of time. Further, the costly read/write operations on HDFS are required for the intermediate result of jobs. Spark keeps the good features of MapReduce, resolves the problems with MapReduce, and adds a number of additional features. The main characteristics of Spark are Speed, rich set of RDD-based operations, in-memory computation, and support to batch, interactive, iterative and streaming data processing.

During the recent years, many Spark-based FIM algorithms have been proposed. Qiu et al. [11] have proposed a Spark-based Apriori algorithm named YAFIM (Yet Another Frequent Itemset Mining). YAFIM is modularized into two phases. The first phase produces frequent items, whereas the second phase generates frequent (k+1)-itemsets from frequent k-itemsets for k ≥ 2. Spark benefits here as data is loaded only one time, and distributed to all workers of the cluster in the form of RDD to utilize the in-memory computation. YAFIM outperformed the MapReduce-based Apriori around 25 times. Rathee et al. [12] proposed R-Apriori (Reduced-Apriori), a parallel Apriori-based algorithm on the Spark RDD framework. R-Apriori is similar to YAFIM with an additional phase that reduces the computation to generate 2-itemsets. The authors considered the time and space complexity of generation of candidate pairs in $2^{nd}$ iteration of Apriori. Complexity is reduced by removing candidate generation, and using a bloom filter instead of a hash tree. Adaptive-Miner [13] is an improvement over the R-Apriori, which dynamically selects a conventional or reduced approach of candidate generation, based on the number of frequent itemsets in recent iteration. Actually, the reduced approach of R-Apriori is not suitable for all iterations, especially when number of frequent itemsets in the previous iteration is less, so conditioning is applied to select appropriate approach. Yang et al. [14] improved the classical Apriori and parallelize it on Spark. The improved Apriori optimizes the pruning step, and avoids the repeated scans of database by mapping it to memory through the array vectors data structure. The Spark version of this algorithm consists of two phases: local frequent itemset generation, and global frequent itemset generation. In the first phase, improved Apriori runs locally on each horizontal partition of database distributed to different workers, whereas the second phase merges all local itemsets and generates global frequent itemsets. DFIMA (Distributed Frequent Itemset Mining Algorithm) [15] is also an Apriori-based algorithm on Spark. It uses a matrix-based pruning approach to reduce the number of candidate itemsets. In the first step, it creates a Boolean vector for each frequent item, and then 2-itemset matrix from Boolean vectors. In the second step, it generates all frequent (k+1)-itemsets from frequent k-itemsets, for k ≥ 2. HFIM (Hybrid Frequent Itemset Mining) [16] exploits the vertical format of the dataset with Apriori algorithm. The smaller size of vertical dataset reduces the cost of dataset scanning. The first phase of the algorithm produces vertical dataset that contains only frequent items. Also, a revised horizontal dataset is obtained by removing infrequent items from the original dataset. Horizontal dataset is distributed on all worker nodes while vertical dataset is shared. In each iteration, the candidates are generated from each transaction to reduce the number of candidates. Shi et al. [17] proposed DFPS (Distributed FP-growth Algorithm based on Spark) algorithm. The first step of the algorithm calculates frequent items from RDD of transactions. The second step repartitions the conditional pattern base, and the third step generates frequent itemsets in parallel from the independent partitions.

## 4. RDD-Eclat Algorithms

We parallelize Eclat algorithm on Spark RDD framework and named it as RDD-Eclat. We propose five slightly different variants of RDD-Eclat by successively applying different strategies and heuristics. EclatV1 is the first implementation, and its successors EclatV2, EclatV3, EclatV4, and EclatV5 are resulted after applying changes in their respective preceding algorithm. All proposed algorithms are modularized into three to four phases. Here phase does not represent a MapReduce phase, but a logical step of the algorithm. Each phase comprises of transformations, actions, and other operations. The notations used in all algorithms of this paper are listed in Table 1 along with their description.

**Table 1:** Some notations used in the algorithms

| Notation | Description |
|---|---|
| tid | Transaction id |
| t | Transaction |
| itemTids | Pair of item and tidset containing that item |
| freqItemTids | Pair of frequent item and tidset containing that item |
| pairList | A list of (key, value) pairs |
| freqItemCounts | Pair of frequent item and its support count |
| freqItemTidsList | List of freqItemTids |
| freqItemsets | Frequent itemsets |
| freqItemList | List of frequent items |
| n | Number of frequent items |
| p | Number of partitions of equivalence classes |
| min_sup | Minimum support count |
| trieL$_1$ | Prefix tree containing frequent items |
| triMatrixMode | Boolean type representing use of triangular matrix optimization |
| triMatrix | Triangular matrix |
| accMatrix | Accumulator variable for triangular matrix |
| freqItemTidsMap | A hashmap containing frequent items as a keys and respective tidsets as values |
| accMap | Accumulator variable for freqItemTidsMap |
| ECList | List of equivalence classes of 1-length prefix along with tidset of members |
| ECs | Equivalence classes |
| EC | Equivalence class |
| EC$_k$ | Equivalence class of k-itemsets |
| L$_{ECk}$ | Frequent itemsets of equivalence class EC$_k$ |

### 4.1 EclatV1

EclatV1 is divided into three phases described as pseudo codes in Algorithm 2, 3, and 4 respectively. Phase-1 (Algorithm 2) takes input as horizontal database and produces output as frequent items with support count, the number of frequent items, and the database in a vertical format for the subsequent use. It first creates an RDD, *transactions* from the database. Here, the database is not partitioned in order to assign a unique transaction identifier, if it is not present in the database. The *flatMapToPair()* transformation maps each transaction to a *(item, tid)* pairs, and creates a paired RDD. The paired RDD is an RDD containing the *(key, value)* pairs. The *groupByKey()* transformation groups all pairs with the same key. The support count of an item is the size of tidset containing the item. The *filter()* transformation removes the items having support count less than *min_sup*, and produces a paired RDD, *freqItemTids* that contains only frequent items along with tidset. The paired RDD, *freqItemCounts* contains *(item, count)* pairs, where count is the support count of item. Here, *(itemTid._1, itemTid._2)* is a *(key, value)* pair of a Tuple2 [23] type object, *itemTid*. Finally, the action, *collect()* returns the entire content of RDD, *freqItemTids* to the driver program where it is sorted in ascending order of support of items and stored in a list. Fig. 1 shows the lineage graph for RDDs in Phase-1 of EclatV1.

**Algorithm 2:** Phase-1 of EclatV1
1:  RDD transactions = sc.textFile("database", 1);
2:  PairRDD itemTids = transactions.flatMapToPair(t -> {
3:     tid = 1;
4:     for each item of t.split(" ")
5:        pairList.add((item, tid));
6:     tid++;
7:     return pairList;
8:  }).groupByKey(s);
9:  PairRDD freqItemTids = itemTids.filter(itemTid -> itemTid._2.size () >= min_sup);
10: PairRDD freqItemCounts = freqItemTids.mapToPair(itemTid -> (itemTid._1, itemTid._2.size(s)));
11: freqItemCounts.saveAsTextFile("frequentItems");
12: freqItemTidsList = sort(freqItemTids.collect());
13: n = freqItemTidsList.size(s);

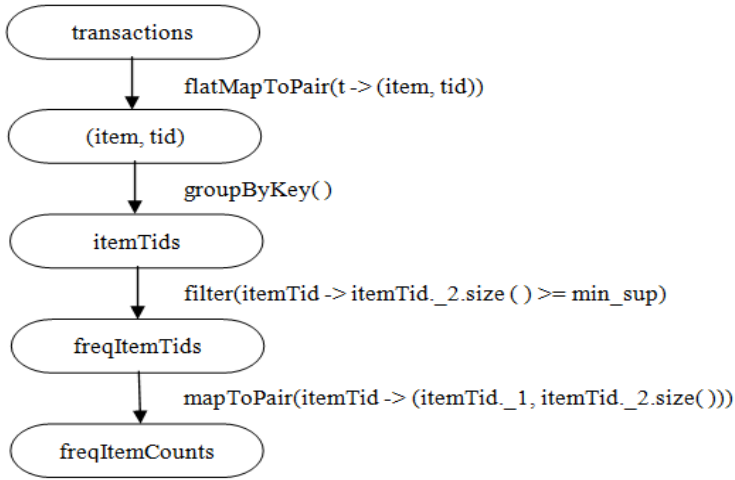

**Fig. 1:** Lineage graph for RDDs in Phase-1 of EclatV1

Phase-2 of EclatV1 (Algorithm 3) computes support count of all 2-itemsets using an upper triangular matrix from the horizontal database, as recommended by Zaki in [3]. It is computed in parallel on the different partitions of the database. The database is partitioned as per the default parallelism which value is equal to the number of cores on all machines of the Spark cluster. The triangular matrix method may or may not be applied depending on the user's choice. The size of matrix depends on the maximum integer value of all items, and for a very large integer it may increase the memory requirement. The triangular matrix is shared as an accumulator variable, *accMatrix* among all executors to add support count of 2-itemsets in parallel. The transformation, *flatMap()* updates the accumulated matrix for all 2-itemset combination of each transaction. The lineage graph for RDDs in Phase-2 of EclatV1 is shown in Fig. 2.

**Algorithm 3:** Phase-2 of EclatV1
1:  transactions = transactions.repartition(sc.defaultParallelism());
2:  if(triMatrixMode)
3:  {
4:     create a triangular matrix, triMatrix[ ][ ]
5:     pass triMatrix as accumulator variable, accMatrix
6:     transactions.flatMap(t -> {
7:        for each 2-itemset combination, itemIitemJ of items of t.split(" ")
8:           accMatrix.update(itemIitemJ);
9:     });
10:    triMatrix = accMatrix.value();
11: }

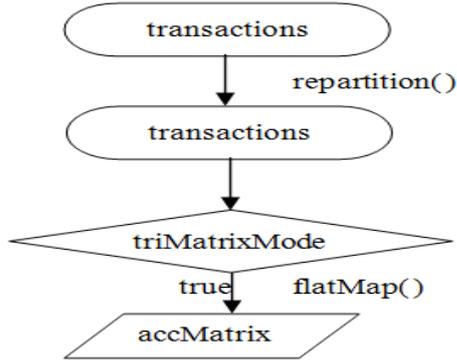

**Fig. 2:** Lineage graph for RDDs in Phase-2 of EclatV1

Phase-3 of EclatV1 (Algorithm 4) takes input as *freqItemTidsList*, the vertical dataset as a list of pairs of frequent items and corresponding tidset, and produces frequent k-itemsets, k ≥ 2. It first generates *ECList*, a list of pairs of equivalence classes for 2-itemsets and tidset of members of the class. A paired RDD, *ECs* is created by parallelizing the *ECList*, and partitioned into default *(n-1)* partitions, where *n* is the number of frequent items. The triangular matrix containing the support count of 2-itemsets is used here to avoid the costly intersection operations for infrequent 2-itemsets. The transformation, *flatMap()* processes each partition of the equivalence classes in parallel. It applies the *Bottom-Up()* recursive function of Eclat (Algorithm 1) on each equivalence class in a partition. The source code of *Bottom-Up()* method has been taken from the SPMF Open-Source Data Mining Library [38]. Fig. 3 shows the lineage graph for RDDs in Phase-3 of EclatV1.

| **Algorithm 4:** Phase-3 of EclatV1 |
|---|
| 1:  for(i = 0; i < freqItemTidsList.size() - 1; i++) |
| 2:  { |
| 3:      itemI = freqItemTidsList.get(i)._1; |
| 4:      tidsetI = freqItemTidsList.get(i)._2; |
| 5:      for(j = i + 1; j < freqItemTidsList.size(); j++) |
| 6:      { |
| 7:          itemJ = freqItemTidsList.get(j)._1; |
| 8:          if(triMatrixMode) |
| 9:              if(triMatrix.getSupport(itemI, itemJ) < min_sup) |
| 10:                 continue; |
| 11:         tidsetJ = freqItemTidsList.get(j)._2; |
| 12:         tidsetIJ = tidsetI ∩ tidsetJ; |
| 13:         prefixIList.add((itemJ, tidsetIJ)); |
| 14:     } |
| 15:     ECList.add(itemI, prefixIList); |
| 16: } |
| 17: PairRDD ECs = sc.parallelize(ECList); |
| 18: ECs = ECs.partitionBy(new defaultPartitioner(n - 1)).cache(s); |
| 19: RDD freqItemsets = ECs.flatMap(EC -> Bottom-Up(EC)); |
| 20: freqItemsets.saveAsTextFile("frequentItemsets"); |

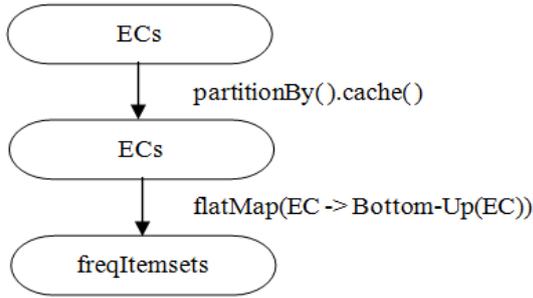

**Fig. 3:** Lineage graph for RDDs in Phase-3 of EclatV1

### 4.2 EclatV2

EclatV2 comprises of four phases, the pseudo codes of first three phases are described in Algorithms 5, 6, and 7, whereas Phase-4 is same as the Phase-3 of EclatV1 (Algorithm 4). EclatV2 applies the filtered transaction technique adopted from [18] to reduce the size of horizontal database. The filtered transaction technique removes the infrequent items from transactions. Consequently, it reduces the cost of operations further applied over the horizontal database, and memory requirement. This optimization is achieved at the cost of additional scanning of database that may amortize if the database is reduced significantly after the filtering. The Phase-1 starts with the partitioning of transactions of the database and creating RDD. The number of partitions is the default that is equal to the number of cores in all machines of the cluster. The transformation, *flatMap()* splits each transaction into items constituting the transaction, then *mapToPair()* converts each item into *(key, value)* pairs, where *key* is the item and *value* is 1. The *reduceByKey()* transformation sums up all values having the same key, then *filter()* transformation checks the sum of values against *min_sup* and returns only frequent items and their support count. Finally, frequent items are returned by the action, *collect()* to the driver, where it is sorted in alphanumeric order, and stored as a list. Here, the level of parallelism is same in all transformations as the number of partitions of parent RDD is preserved throughout the lineage. Fig. 4 shows the lineage graph for RDDs in Phase-1 of EclatV2.

**Algorithm 5:** Phase-1 of EclatV2
1: RDD transactions = sc.textFile ("database");
2: RDD items = transactions.flatMap(t -> List(t.split(" ")));
3: PairRDD itemPairs = items.mapToPair(item -> (item, 1));
4: PairRDD itemCounts = itemPairs.reduceByKey(($v_1$, $v_2$) -> $v_1$ + $v_2$);
5: PairRDD freqItemCounts = itemCounts.filter(itemCount -> itemCount._2 >= min_sup);
6: freqItemCounts.saveAsTextFile("frequentItems");
7: freqItemList = sort(freqItemCounts.keys().collect());
8: n = freqItemList.size();

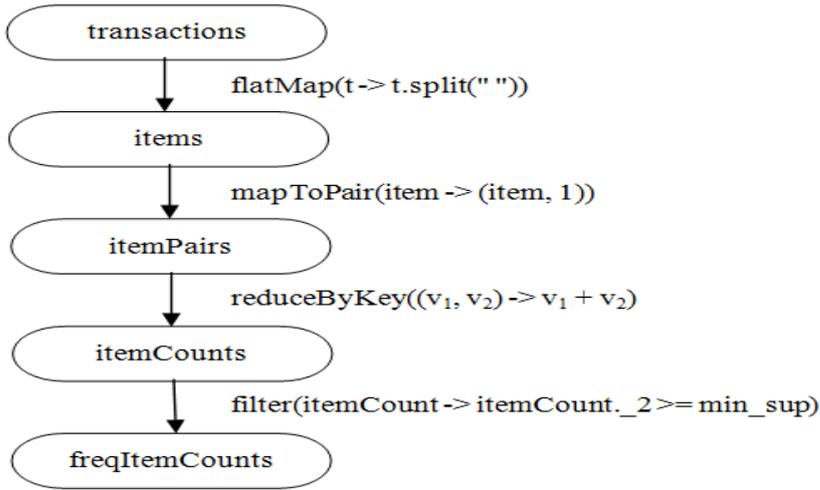

**Fig. 4:** Lineage graph for RDDs in Phase-1 of EclatV2

Phase-2 of EclatV2 (Algorithm 6) is similar to the Phase-2 of EclatV1 except the addition of transaction filtering. First, the *map()* transformation applies the *filterTransaction()* method for each transaction in parallel and returns filtered transactions. The frequent items, *trieL$_1$* stored in a prefix tree, must be broadcasted to all executors using the broadcast variable, before applying the transformation. The support counting of 2-itemsets is performed applying the same triangular matrix method of EclatV1, but on the filtered transactions. The lineage graph of this phase is shown in Fig. 5.

**Algorithm 6:** Phase-2 of EclatV2
1:   store frequent items in trie, trieL$_1$;
2:   RDD filteredTransactions = transactions.map(t -> filterTransaction(t.split(" "), trieL$_1$));
3:   if(triMatrixMode)
4:   {
5:      create a triangular matrix, triMatrix[ ][ ]
6:      pass triMatrix as accumulator variable, accMatrix
7:      filteredTransactions.flatMap(t -> {
8:        for each 2-itemset combination, itemIitemJ of items of t.split(" ")
9:           accMatrix.update(itemIitemJ);
10:     });
11:     triMatrix = accMatrix.value();
12: }

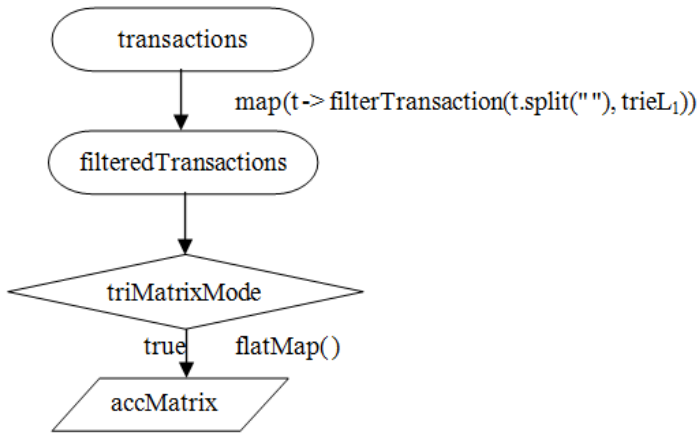

**Fig. 5:** Lineage graph for RDDs in Phase-2 of EclatV2

Phase-3 of EclatV2 (Algorithm 7) generates the vertical dataset from filtered horizontal dataset. It first reduces all partitions of transactions into one partition in order to generate unique transaction identifier. The *flatMapToPair()* transformation generates *(item, tid)* pairs for each transaction, then *groupByKey()* generates *(item, tidset)* pairs by grouping tid's of same frequent item. Frequent items are already computed in Phase-1, so there is no need of further transformations. The action *collect()* returns the list of *(item, tidset)* pairs, that is sorted by the total order of increasing support count and stored in a list. Fig. 6 shows the lineage graph of this phase. Phase-4 of EclatV2 is exactly same as the Algorithm 4, where equivalence classes are created and partitioned for parallel computation of the frequent itemsets.

**Algorithm 7:** Phase-3 of EclatV2
```
1:  filteredTransactions = filteredTransactions.coalesce(1);
2:  PairRDD freqItemTids = filteredTransactions.flatMapToPair(t -> {
3:      tid = 1;
4:      for each item of t.split(" ")
5:          pairList.add((item, tid));
6:      tid++;
7:      return pairList;
8:  }).groupByKey( );
9:  freqItemTidsList = sort(freqItemTids.collect( ));
```

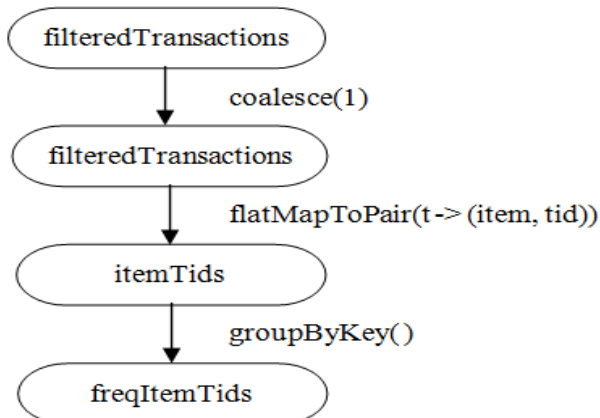

**Fig. 6:** Lineage graph for RDDs in Phase-3 of EclatV2

### 4.3 EclatV3

EclatV3 comprises of four phases in which first two phases, Phase-1 and Phase-2 are exactly same as those of EclatV2. The purpose of Phase-3 of both algorithms EclatV2 and EclatV3 is same i.e. both generate vertical dataset. The difference is that EclatV3 uses a hashmap data structure to store *(item, tidset)* pairs of vertical dataset. This hashmap is accumulated across all executors, and updated by the *flatMapToPair()* transformation. The hashmap, *freqItemTidsMap* is used to sort the list of frequent items of Phase-1, by total order of increasing support count. Algorithm 8 describes the pseudo code of Phase-3, and Fig. 7 shows the lineage graph for RDDs.

**Algorithm 8:** Phase-3 of EclatV3
1: filteredTransactions = filteredTransactions.coalesce(1);
2: create a hashmap, freqItemTidsMap
3: pass freqItemTidsMap as accumulator variable, accMap
4: filteredTransactions.flatMapToPair(t -> {
5:   tid = 1;
6:   accMap.update((t, tid));
7:   tid++;
8: });
9: freqItemTidsMap = accMap.value(s);
10: freqItemList = sort(freqItemList, freqItemTidsMap);

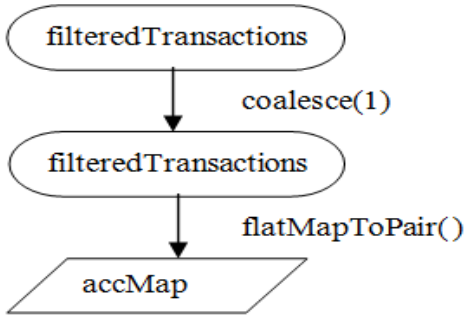

**Fig. 7:** Lineage graph for RDDs in Phase-3 of EclatV3

Phase-4 of EclatV3 (Algorithm 9) is similar to the Algorithm 4, the only difference is the data structure used to store the pairs of item and tidset. Here, the items and corresponding tidsets are fetched from a hashmap, *freqItemTidsMap* instead of a list, and the rest of the process is same. So, the lineage graph for RDDs will be same as in Fig. 3.

**Algorithm 9:** Phase-4 of EclatV3
1: for(i = 0; i < freqItemList.size() - 1; i++)
2: {
3:   itemI = freqItemList.get(i);
4:   tidsetI = freqItemTidsMap.get(itemI);
5:   for(j = i + 1; j < freqItemList.size(); j++)
6:   {
7:     itemJ = freqItemList.get(j);
8:     if(triMatrixMode)
9:       if(triMatrix.getSupport(itemI, itemJ) < min_sup)
10:        continue;
11:    tidsetJ = freqItemTidsMap.get(itemJ);
12:    tidsetIJ = tidsetI ∩ tidsetJ;
13:    prefixIList.add((itemJ, tidsetIJ));
14:  }
15:  ECList.add(itemI, prefixIList);
16: }
17: PairRDD ECs = sc.parallelize(ECList);
18: ECs = ECs.partitionBy(new defaultPartitioner(n - 1)).cache();
19: RDD freqItemsets = ECs.flatMap(EC -> Bottom-Up(EC));
20: freqItemsets.saveAsTextFile("frequentItemsets");

## 4.4 EclatV4 and EclatV5

Algorithms EclatV4 and EclatV5 apply the heuristics on EclatV3 to partition the equivalence classes into $p$ partitions, where $p$ has the value supplied by the user. Heuristics are applied to balance the partitions of equivalence classes. Only the Phase-4 of these two algorithms is different from EclatV3, and first three phases are the same to those of EclatV3. Further, Phase-4 is different only with respect to the partitioning of equivalence classes at line no. 18 (Algorithm 9). EclatV4 and EclatV5 respectively use *hashPartitioner* and *reverseHashPartitioner* in their Phase-4. So, in Phase-4 of EclatV4 and EclatV5, line no. 18 of Algorithm 9 is simply replaced by the following lines respectively.

18: ECs = ECs.partitionBy(new hashPartitioner(p)).cache( );
18: ECs = ECs.partitionBy(new reverseHashPartitioner(p)).cache( );

## 4.5 Equivalence Class Partitioners

In general, an efficient partitioning of RDDs reduces data shuffling across the network in the subsequent transformations. This section describes the partitioning techniques used in the proposed algorithms to partition the RDDs of equivalence classes, *ECs*, for the parallel and independent computation of frequent itemsets. Partitioning is done on the basis of prefixes of equivalence classes. Spark facilitates to implement custom partitioner; our custom partitioners are based on the HashPartitioner, a default partitioner of Spark. A custom partitioner extends the Partitioner class of Spark, and implements its *getPartition()* method. The heuristic of partitioning is defined in this method. Algorithm 10 describes the pseudo codes of *getPartition(v)* method of three custom partitioners, where $v$ is the unique value assigned to the 1-length prefix of equivalence classes. A hash map is used to map each frequent item to a unique integer between *0* and *n-1*, where *n* is number of frequent items. EclatV1, EclatV2, and EclatV3 create partitions individually for each equivalence class, i.e. *(n-1)* partitions. It is termed as default partitioning. EclatV4 and EclatV5 limit the number of partitions, and create $p$ partitions, where the value of $p$ is supplied by the user. The numbers of partitions determine the number of parallel tasks. The right number partition is an important factor for better performance, and it should not be too many and too few. The partitioner used by EclatV4 is termed as hash partitioner that applies a hash function on the values corresponding to the prefix of equivalence classes. It simply divides the value, $v$ by $p$, and returns the remainder as a partition ID. EclatV5 also creates $p$ partitions of equivalence classes, but returns the partition ID in reverse order when $v \geq p$. The partitioners with hashing and reverse hashing are used to investigate the workload balance among partitions. The workload is measured in terms of the members in equivalence classes. An equivalence class having more members leads to the generation of more candidate itemsets as well as the intersection of their tidsets.

**Algorithm 10:** Equivalence Class Partitioners

```
// getPartition method of defaultPartitioner
1:   getPartition(v)
2:   {
3:       return v;
4:   }
// getPartition method of hashPartitioner
1:   getPartition(v)
2:   {
3:       return v % p;
4:   }
// getPartition method of reverseHashPartitioner
1:   getPartition(v)
2:   {
3:       r = v % p;
4:       if(v >= p)
5:           return (p - 1) - r;
6:       else
7:           return r;
8:   }
```

## 5. Experimental Results

This section presents the experimental environment, used datasets, and performance of the algorithms in terms of execution time. Performance analysis is presented in three categories. First, performance of the proposed algorithms is compared with the Spark-based Apriori implementation similar to YAFIM [11], on the different datasets for the varying value of minimum support. Further, performance of the proposed algorithms is also compared with each other on the different datasets for the varying value of minimum support. Second, performance of the proposed algorithms is analyzed on the different datasets for the increasing number of executor cores; and the third one is scalability test on the increasing size of dataset.

### 5.1 Experimental Setup and Datasets

The Experiments are conducted on a workstation machine installed with Spark-2.1.1, Hadoop-2.6.0, and Scala-2.11.8. The workstation is equipped with Intel Xenon CPU E5-2620@2.10 GHz with 24 cores, 16 GB memory and 1 TB disk, and running 64 bit Ubuntu 14.04. HDFS is used as storage for the input datasets and generated frequent itemsets. Source codes of all algorithms are written in Java-7.

Algorithms are evaluated on the seven benchmark datasets of both type, i.e. synthetics and real life. Table 2 summarizes these datasets. The datasets c20d10k, BMS_WebView_1, and BMS_WebView_2 (in short BMS1 and BMS2) are taken from SPMF datasets [38], whereas the datasets chess, mushroom, T10I4D100K, and T40I10D100K are taken from the FIMI dataset repository [39].

**Table 2:** Datasets used in experiments with their properties

| Dataset | Type of dataset | Transactions | Items | Average Transaction Width |
|---|---|---|---|---|
| c20d10k | Synthetic | 10,000 | 192 | 20 |
| chess | Real-life | 3196 | 75 | 37 |
| mushroom | Real-life | 8124 | 119 | 23 |
| BMS_WebView_1 | Real-life | 59602 | 497 | 2.5 |
| BMS_WebView_2 | Real-life | 77512 | 3340 | 5 |
| T10I4D100K | Synthetic | 100,000 | 870 | 10 |
| T40I10D100K | Synthetic | 100,000 | 1000 | 40 |

### 5.2 Performance Analysis

The proposed algorithms EclatV1, EclatV2, EclatV3, EclatV4, and EclatV5 require two parameters, *triMatrixMode* and *p* to be set before the execution. The triangular matrix optimization can be enabled or disabled by providing the true or false value to *triMatrixMode*. It is true for all datasets except BMS1 and BMS2. The size of triangular matrix depends on the maximum integer value of all items in the dataset, and it is very large for datasets BMS1 and BMS2. The very large size of the matrix may cause out of memory problem, so the value of *triMatrixMode* is false for these two datasets. Further, algorithms EclatV4 and EclatV5 partition the equivalence classes into *p* partitions, where the value of *p* is specified by the user. For the all datasets, the value of *p* is set to 10.

*5.2.1 Execution Time on Varying Value of Minimum Support*

All algorithms are compared with respect to the execution time on the varying value of minimum support for all datasets summarized in Table 2. Execution times of the proposed RDD-Eclat based algorithms are compared with each other as well as with the Apriori algorithm. Figs. 8-14 show the execution time of various algorithms on the datasets of Table 2 for the varying value of minimum support. Figs. 8(a)-14(a) compare the execution time of the proposed algorithms against the Apriori algorithm whereas Figs. 8(b)-14(b) compare the execution time of the proposed algorithms EclatV1, EclatV2, EclatV3, EclatV4, and EclatV5. On all datasets, RDD-Eclat outperforms the RDD-Apriori (Figs. 8(a)-14(a)), and the execution time difference between them becomes wider with the decreasing value of minimum support. If we consider Eclat1 to compare with Apriori, then it can be seen that EclatV1 is at least nine times faster than Apriori on the datasets BMS2 and T40I10D100K (Figs. 12(a) and 14(a)), seven times on the dataset chess (Fig. 9(a)), six times on the dataset BMS1 (Fig. 11(a)), four times on the dataset c20d10k, and two times on the datasets mushroom and T10I4D100K (Figs. 10(a) and 13(a)), at the lowest value of minimum support in each case. In short, RDD-Eclat outperforms Apriori by at least two times and up to nine times on some datasets.

A number of heuristics have been applied over EclatV1 in order to further improve its performance. As discussed in section 4 that EclatV1 is the first implementation of the RDD-Eclat, and its successors EclatV2, EclatV3, EclatV4, and EclatV5 are resulted after applying changes in their respective preceding algorithms. Since, the Apriori algorithm is outperformed by these proposed algorithms, all the subsequent observations are considered only for the proposed algorithms. Figs. 8(b)-14(b) closely compare the execution time of the proposed algorithms EclatV1, EclatV2, EclatV3, EclatV4, and EclatV5. The major algorithmic difference between EclatV1 and EclatV2, EclatV3 is the use of filtered transaction technique in EclatV2 and EclatV3; and the difference between EclatV2, EclatV3 and EclatV4, EclatV5 is the use of hash partitioners for the equivalence class partitioning. The performance improvement of EclatV2 and EclatV3 over EclatV1 is not significant (Figs. 8(b)-10(b)); even they perform worse than EclatV1 on some datasets (Figs. 11(b)-14(b)). Algorithms EclatV2 and EclatV3 only improve the performance when they significantly reduce the size of the original transactions after applying the filtered transaction technique. If the size of filtered transactions is still near to that of the original transactions, then it only adds overhead, and increases the overall execution time of the algorithms. For example, for the dataset T40I10D100K, the resulting filtered transaction is reduced only by 3.2 %, 8.4 %, 16.1 %, and 25.8 % on the minimum support 0.01, 0.02, 0.03, and 0.04 respectively. Adoption of the filtered transaction technique may improve the performance on a dataset of larger scale where the filtered dataset is reduced significantly. Further, it can be seen that algorithms EclatV4 and EclatV5 significantly improve the performance in comparison to EclatV2 and EclatV3 on all datasets (Figs. 8(b)-14(b)). It proves the effectiveness of equivalence class partitioners used in EclatV4 and EclatV5.

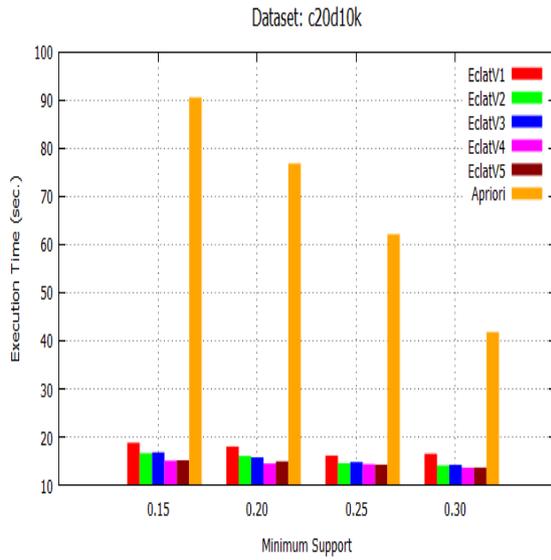
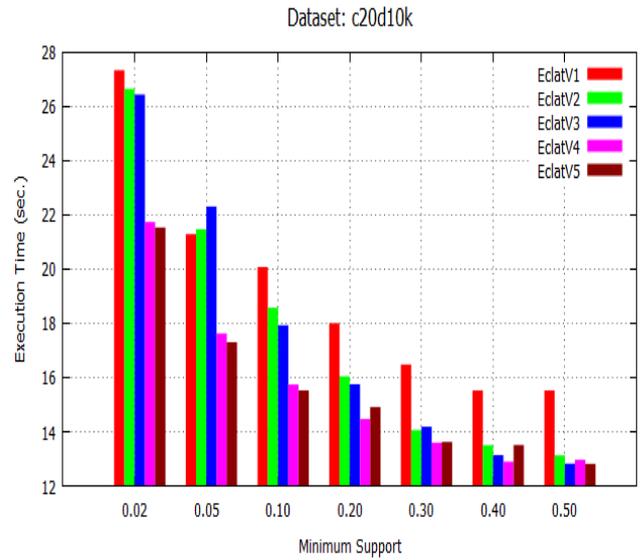

**Fig. 8:** Execution time of algorithms (a) EclatV1, EclatV2, EclatV3, EclatV4, EclatV5, and Apriori (b) EclatV1, EclatV2, EclatV3, EclatV4, and EclatV5 for varying minimum support on dataset c20d10k.

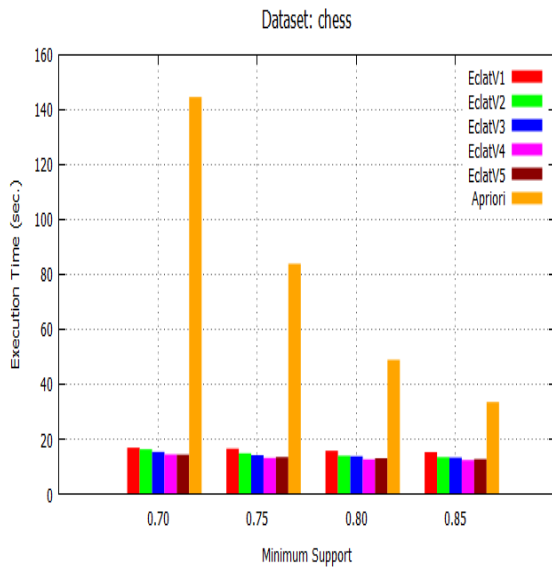
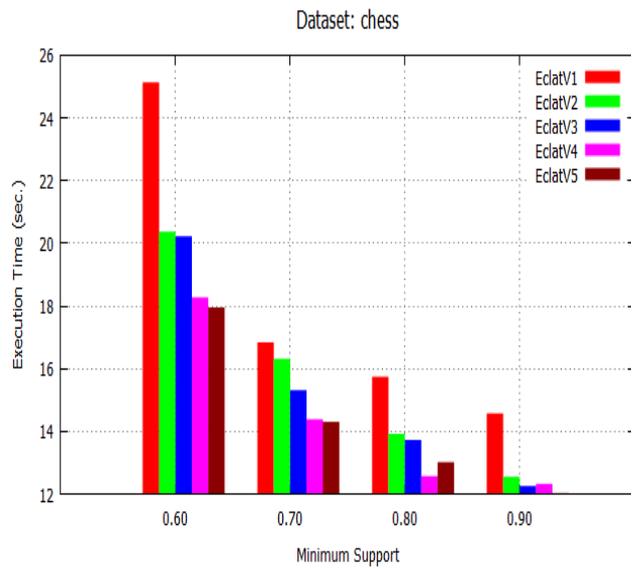

**Fig. 9:** Execution time of algorithms (a) EclatV1, EclatV2, EclatV3, EclatV4, EclatV5, and Apriori (b) EclatV1, EclatV2, EclatV3, EclatV4, and EclatV5 for varying minimum support on dataset chess.

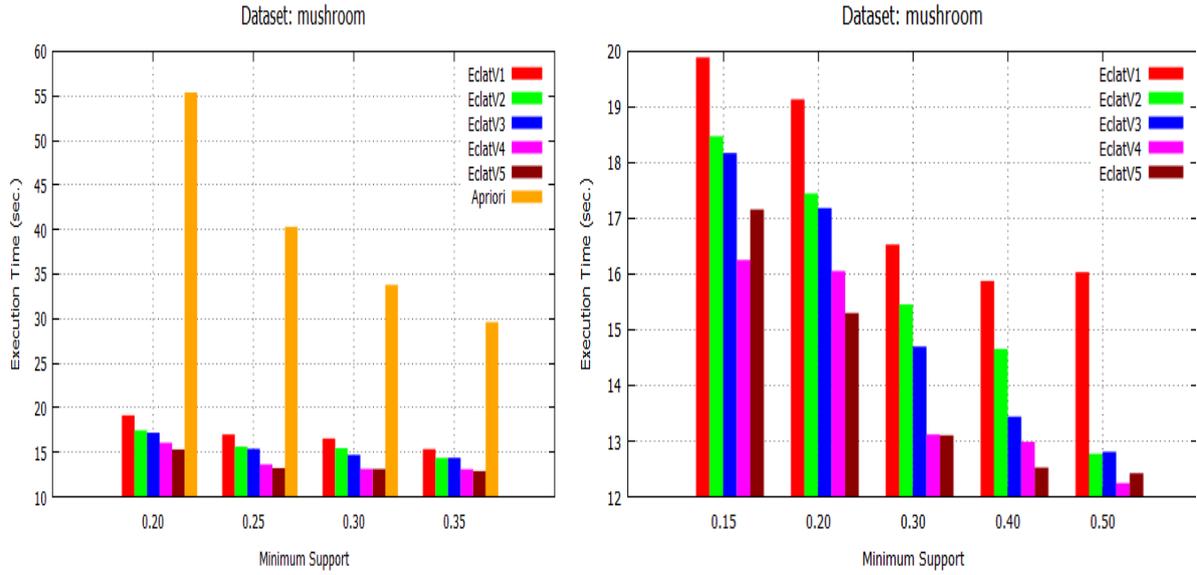

**Fig. 10:** Execution time of algorithms (a) EclatV1, EclatV2, EclatV3, EclatV4, EclatV5, and Apriori (b) EclatV1, EclatV2, EclatV3, EclatV4, and EclatV5 for varying minimum support on dataset mushroom.

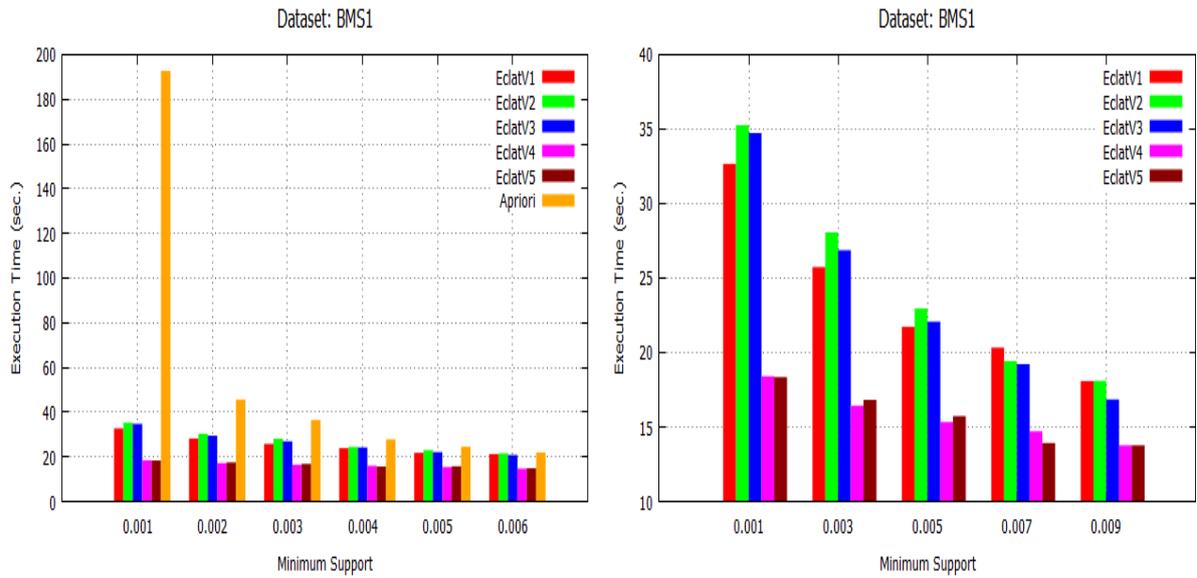

**Fig. 11:** Execution time of algorithms (a) EclatV1, EclatV2, EclatV3, EclatV4, EclatV5, and Apriori (b) EclatV1, EclatV2, EclatV3, EclatV4, and EclatV5 for varying minimum support on dataset BMS_WebView_1.

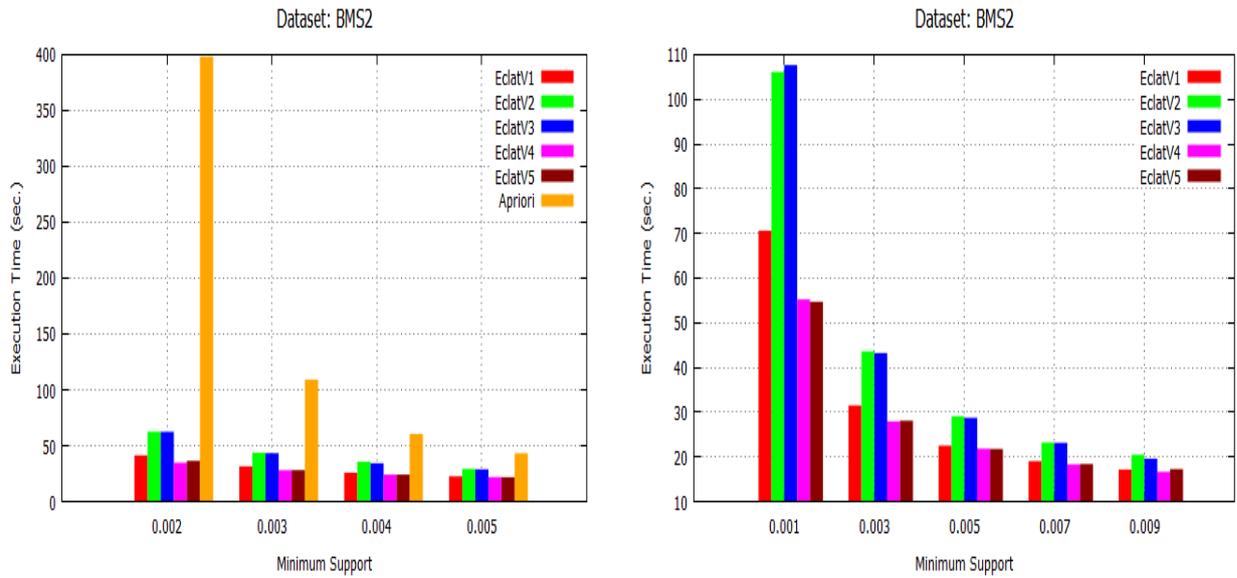

**Fig. 12:** Execution time of algorithms (a) EclatV1, EclatV2, EclatV3, EclatV4, EclatV5, and Apriori (b) EclatV1, EclatV2, EclatV3, EclatV4, and EclatV5 for varying minimum support on dataset BMS_WebView_2.

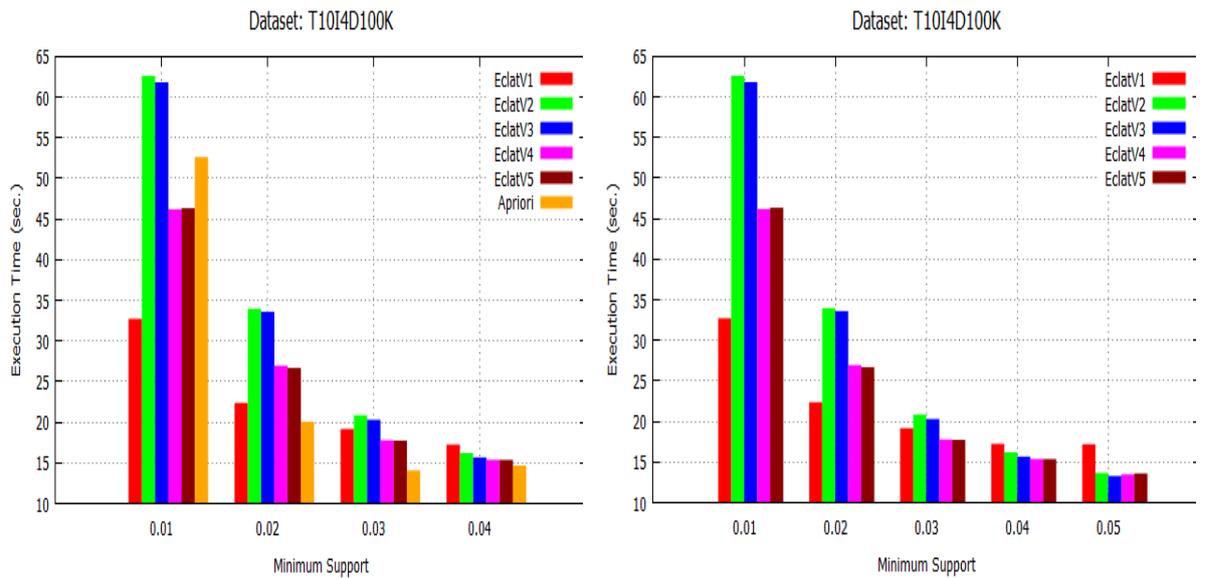

**Fig. 13:** Execution time of algorithms (a) EclatV1, EclatV2, EclatV3, EclatV4, EclatV5, and Apriori (b) EclatV1, EclatV2, EclatV3, EclatV4, and EclatV5 for varying minimum support on dataset T10I4D100K.

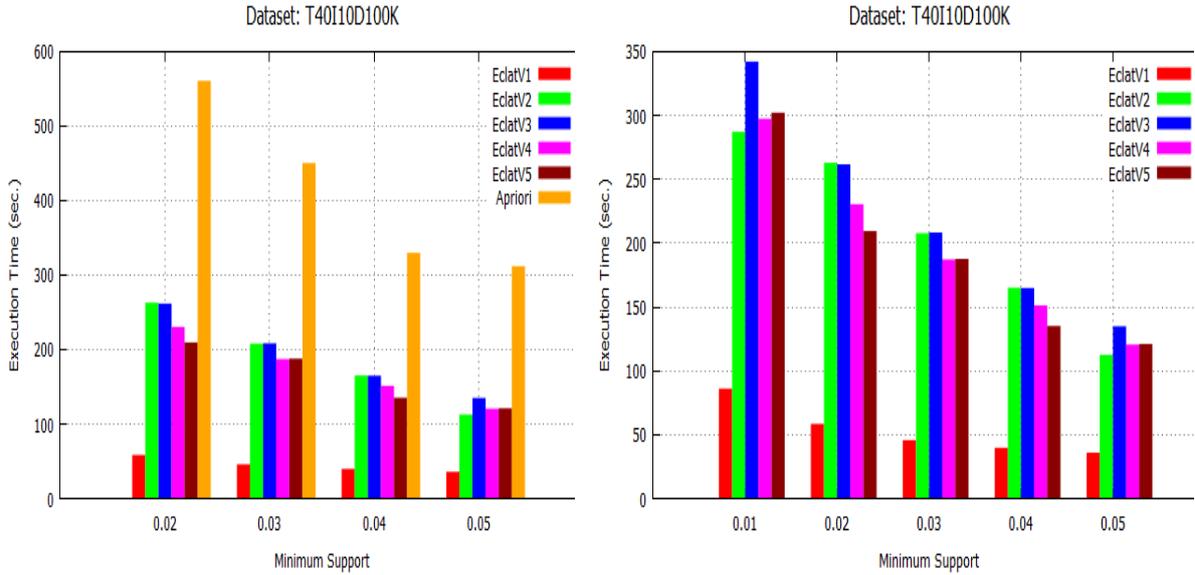

**Fig. 14:** Execution time of algorithms (a) EclatV1, EclatV2, EclatV3, EclatV4, EclatV5, and Apriori (b) EclatV1, EclatV2, EclatV3, EclatV4, and EclatV5 for varying minimum support on dataset T40I10D100K.

*5.2.2 Execution Time on Increasing Number of Executor Cores*

The behavior of the proposed algorithms is investigated on the different datasets for the increasing number of executor cores, as shown in Fig. 15(a-e). Execution time has been measured using 2, 4, 6, 8, and 10 executor cores for the five datasets. With the increasing number of cores, execution time of the algorithms decreases. The decline is more apparent in the algorithms taking more time than those taking comparatively less time (Fig. 15(a-e)). It indicates that execution time can be reduced or maintained by allocating more cores or by adding more nodes.

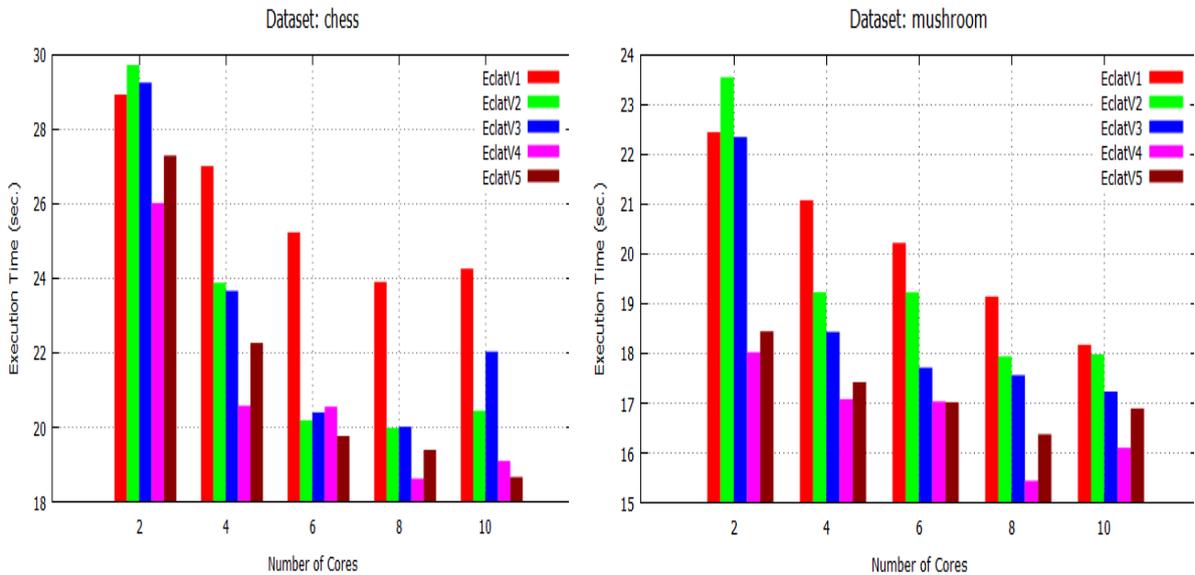

(a) Dataset chess at min_sup = 0.60         (b) Dataset mushroom at min_sup = 0.15

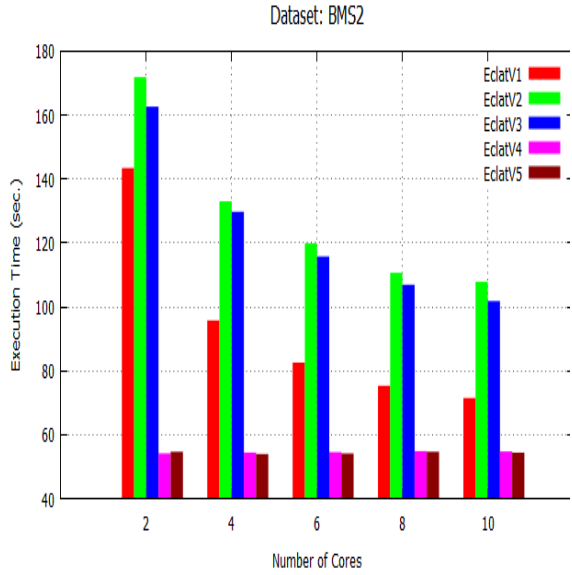

(c) Dataset BMS_WebView_2 at min_sup = 0.001

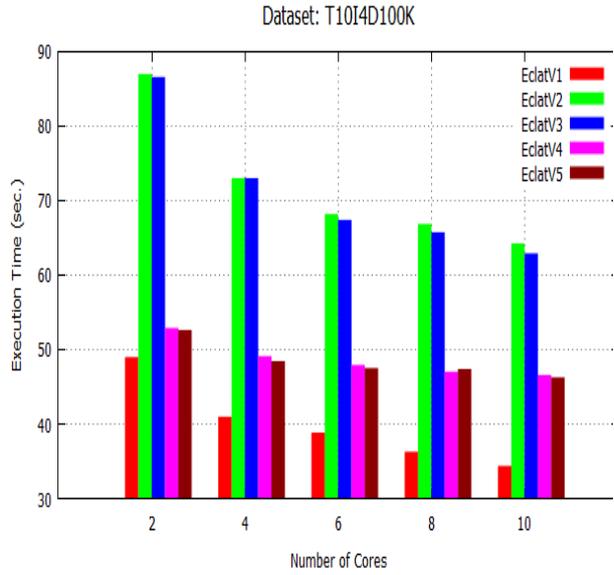

(d) Dataset T10I4D100K at min_sup = 0.01

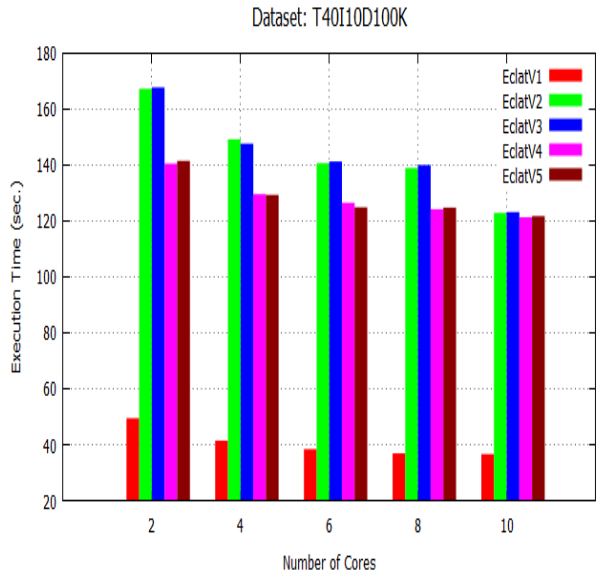

(e) Dataset T40I10D100K at min_sup = 0.01

**Fig. 15:** Execution time on varying number of executor cores for five datasets

*5.2.3 Scalability on Increasing Size of Dataset*

Scalability test is carried out for the proposed algorithms on the increasing size of dataset T10I4D100K at a fixed value of minimum support, 0.05. To get the larger dataset size, it is doubled each time from its previous dataset, so it ranges from 100K transactions to 1600K transactions as shown in Fig. 16. It can be seen in the figure that with the increasing dataset size, execution time of all algorithms increases linearly.

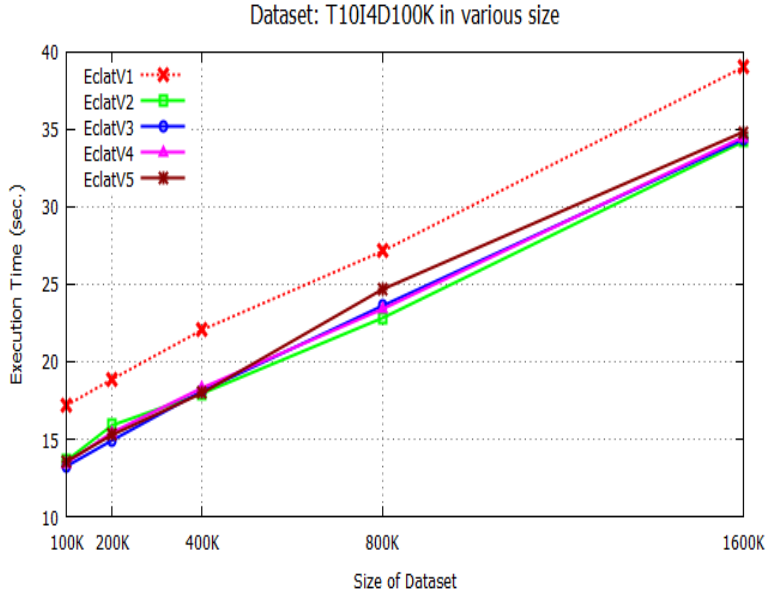

**Fig. 16:** Execution time on increasing size of dataset T10I4D100K at min_sup = 0.05

## 6. Conclusions and Future Directions

An unexplored problem, re-designing Eclat algorithm in the distributed computing environment of Spark, has been explored in this paper. The key contribution here is a parallel Eclat algorithm on the Spark RDD framework, named as RDD-Eclat along with the implementation of its five variants. The first variant is EclatV1, and the subsequent variants are EclatV2, EclatV3, EclatV4, and EclatV5. Each variant is resulted from applying some different approach and heuristic on the previous variant. The filtered transaction technique is applied after EclatV1, and the heuristics for equivalence class partitioning are applied in EclatV4 and EclatV5. Algorithmically, EclatV2 and EclatV3 are slightly different as well as EclatV4 and EclatV5. Experimental results on the both synthetic and real life datasets, shows that all proposed algorithms outperform the YAFIM, a Spark-based Apriori algorithm, by many times, in terms of execution time. As the minimum support threshold decreases, the proposed algorithms perform better in comparison to Spark-based Apriori. Further, the proposed algorithms are closely compared in order to investigate the effect of various heuristics applied on the latter variants EclatV2, EclatV3, EclatV4, and EclatV5. It has been observed that the filtered transaction technique improves the performance when it significantly reduces the size of the original dataset. Further, the heuristics applied in equivalence class partitioning significantly reduce the execution time. Also, the algorithms exhibit scalability when executed on increasing the number of cores and the size of dataset.

Moreover, a more optimized and fine-tuned RDD-Eclat algorithm can be designed in future by efficiently assembling the different modules from the different variants. For example, the heuristic of equivalence class partitioning is not applied in EclatV1 but in EclatV4 and EclatV5 along with the filtered transaction technique. This paper only considers 1-length prefix based equivalence classes, results can be explored for the k-length prefixes where $k \geq 2$. Also, the heuristic for equivalence class partitioning can be improved further to get a more balanced distribution of equivalence classes.